\documentclass[superscriptaddress, amsmath,amssymb,twocolumn]{revtex4-1}

\usepackage{epsfig,amssymb,amsmath,amsthm,amsfonts,amsbsy,mathrsfs}
\usepackage{graphicx}
\usepackage{color}
\usepackage{bm}% bold math
\usepackage{multirow}

\newcommand{\mX}{\mathcal{X}}

\newcommand{\mD}{\mathcal{D}}

\newcommand{\mH}{\mathcal{H}}
\newcommand{\mbH}{\mathbb{H}}

\newcommand{\be}{\begin{equation}}
\newcommand{\ee}{\end{equation}}

\newcommand{\bea}{\begin{eqnarray}}
\newcommand{\eea}{\end{eqnarray}}

\newcommand{\matteo}[1]{\textcolor{black}{#1}}

\newcommand{\gcol}[1]{\textcolor{black}{#1}}

\usepackage{tikz}
\usetikzlibrary{arrows,shapes,chains}

\begin{document}
\preprint{APS/123-QED} \title{ How non-equilibrium correlations in active matter\\ reveal
  the topological crossover in glasses}

%%%%%%%%%%%%%%%%%%%%%%%%%%%%%%%%%%%%%%%%%%%%%%%%%%%%%%%%%%%%%%%%%%%%%%%%%%%%%
\author{Giacomo Gradenigo}
\email{ giacomo.gradenigo@gssi.it}
\affiliation{Gran Sasso Science Institute, Viale F. Crispi 7, 67100 L’Aquila, Italy}
\affiliation{INFN-Laboratori Nazionali del Gran Sasso, Via G. Acitelli 22, 67100 Assergi (AQ), Italy}

\author{Matteo Paoluzzi}
\email{matteopaoluzzi@ub.edu}
\affiliation{Departament de Física de la Mat\`eria Condensada, Universitat de Barcelona, C. Martí Franqu\`es 1, 08028 Barcelona, Spain.}

\date{\today}% It is always \today, today,

\begin{abstract}
As shown by early studies on mean-field models of the glass
transition, the geometrical features of the energy landscape provide
fundamental information on the dynamical transition at the
Mode-Coupling temperature $T_d$. We show that active particles can
serve as a useful tool for gaining insight into the topological
crossover in model glass-formers. In such systems the landmark of the
minima-to-saddle transition in the potential energy landscape, taking
place in the proximity of $T_d$, is the critical slowing down of
dynamics. Nevertheless, the critical slowing down is a bottleneck for
numerical simulations and the possibility to take advantage of the new
smart algorithms capable to thermalize down in the glass phase is
attractive. Our proposal is to consider configurations equilibrated
below $T_d$ and study their dynamics in the presence of a small amount
of self-propulsion. As exemplified here from the study of the $p$-spin
model, the presence of self-propulsion gives rise to critical
off-equilibrium \emph{equal time} correlations at the
minima-to-saddles crossover, correlations which are not hindered by
the sluggish glassy dynamics.
\end{abstract}

\maketitle

\section{Introduction}
\label{sec:intro}

The detection of order in glass-forming systems has always been
elusive. The first progress was made looking at non-trivial
correlations in dynamical heterogeneities more than twenty years ago~\cite{DFGP02}. Shortly after the attention concentrated also on static multipoint correlations~\cite{BB04,CGV07,BBCGV08,BK12,GTCGV13,BCY16}.  According to mean-field-like scenarios, dynamical correlations are expected to be critical at the Mode-Coupling temperature
$T_d$~\cite{BB04b,FM07,CGB13}, whereas static multipoint correlations are expected to diverge at the ideal glass transition temperature $T_K$, where $T_K<T_d$~\cite{BB04,FM07,CGB13}. The behaviour of static
and dynamic length-scales in glass-formers is of great interest. But, at the same time, its precise characterization is plagued by great difficulties: numerical simulations provide only indirect evidence for the existence of dynamical singularities, which are well defined strictly speaking only in the thermodynamic limit. At the same time, the study of multipoint correlations based on theoretically well defined but practically hard to realize protocols~\cite{BB04,CGV07,BBCGV08,GTCGV13,BCY16}: in experiments their existence cannot be measured directly but only inferred from the behaviour of higher-order
susceptibilities~\cite{BBBCEHLP05,ABMBBLLTWL16}. \\

What makes the search for criticality at $T_d$ particularly interesting
is the lack of any sort of configurational order in glasses. \gcol{The amorphous order of glasses, being either static or dynamic, usually requires the comparison of different configurations to be detected. In particular, it cannot be read off from a single snapshot of the system. The only remarkable exception to the above situation is represented by the \emph{patch correlation length}~\cite{KL10,CB12}, which nevertheless requires knowledge of microscopic details of the systems which are usually out of reach. The goal of our proposal is to compensate for this lack of \emph{configurational} information by putting the system slightly out of equilibrium without perturbing too much its landscape. We will show how, gently pushing the system out of equilibrium, one can gain information on the topological transition from stationary non-equilibrium velocity correlations.}\\

For a given configuration of the system, low-energy excitations
provide a useful tool for distinguishing whether a disorder
configuration belongs to a liquid or glass. In particular, linearly 
unstable configurations, characterized by the presence of negative
eigenvalues in the spectrum, arise in the supercooled liquid phase as the
temperature increases above $T_d$. Amorphous solid configurations
below $T_d$ are on the contrary close to local minima of the potential
energy landscape, \gcol{ so that the corresponding eigenvalues of the Hessian matrix are all non-negative. This picture emerges clearly 
from mean-field models~\cite{BBCZG00,C01,CGP01,GCGP02,castellani2005spin}, where the presence of a crossover from a high-temperature saddle-dominated phase to a low-temperature minima-dominated one is an analytical result.}\\

\gcol{The geometrical features of the energy landscape, which are
  known to influence on equal-time velocity
  correlations~\cite{MGPMD16}, play an important role also in Active Matter~\cite{MM15}.} Recent studies show that active particles develop non-trivial interparticle velocity
correlations~\cite{MGPMD16,PhysRevLett.124.078001,caprini2020hidden,caprini2020active} and equal-time velocity correlations have to be taken into account
also for developing a mode-coupling theory of active
particles~\cite{SFB15,S16,NG17}. Those correlations play an important
role in rationalizing the emergence of collective patterns in dense
active systems~\cite{MGPMD16,henkes2020dense}. Quite remarkable are
also the investigations of static multi-points correlations and
amorphous order in dense active systems~\cite{NMBDRG18}.\\

\gcol{The present work aims to test whether the presence of
  active dynamics and the corresponding stationary non-equilibrium
  correlations offer an \emph{extra} tool to detect the
  minima-to-saddles crossover in glass-forming liquids, rather than being
  regarded just as a disturbance to the natural tendency of the system
  towards a glassy arrested
  state~\cite{KSZ10,GFZ11,KSZ13,BK13,B14,SFB15,S16}}.

\section{The potential energy landscape of glass-formers}
\label{sec:one}

\gcol{It is well known that for glass-forming systems the crossover to
  activated relaxation below $T_d$ corresponds to a topological
  crossover in the landscape~\cite{BBCZG00,C01,CGP01,GCGP02}: when
  $T>T_d$ equilibrium configurations are typically close to an
  unstable stationary point of the potential energy while at low
  temperatures, $T<T_d$, equilibrium configurations are typically
  close to a
  minimum~\cite{BBCZG00,C01,CGP01,GCGP02,PhysRevLett.85.5356,angelani2002quasisaddles}.}

In particular, considering a system composed of $N$ particles
interacting via a translational invariant pair potential $U(
| {\bf x}_i - {\bf x}_j | )$ and where the total potential energy is
$\Phi=\sum_{i<j} U(|{\bf x}_i-{\bf x}_j|)$, the stability of a given
configuration depends on the distribution of the eigenvalues of the
Hessian matrix $ \mbH \equiv \partial^2U/\partial x_i^\alpha\partial
x_j^\beta$, where latin indices denotes particles and greek indices
cartesian components in $d=3$ space. \gcol{When all the eigenvalues of
  $\mbH$ are positive, the system is in the minima-dominated region:
  in mean-field models ergodicity is broken dynamically. As soon as a
  fraction of negative eigenvalues appears, relaxation starts to take
  place along the unstable directions of the Hessian. The evidence of
  a topological transition could be in principle obtained from the
  study of the landscape in equilibrium configurations \cite{CGP01}.
  But this is in practice impossible since it requires to know
  microscopic details, in particular particles positions, which are
  usually not accessible in experiments. Our goal is to show that
  stationary non-equilibrium velocity correlations, measurable even at
  a coarse-grained scale, are already carrying information on the
  landscape structure.}

\section{Active dynamics}
\label{sec:two}

\gcol{ We first consider, as the most generic case, an off-lattice
  systems made of $N$ particles interacting with the pair potential
  $U(|{\bf x}_i-{\bf x}_j|)$. Since the model system we have in mind
  is a colloidal glass or, more in general, a dense active system at
  low Reynolds numbers, we consider overdamped equation of
  motions~\cite{Marchetti13}. The choice between overdamped or
  inertial dynamics is of course in general not arbitrary, it depends
  on the system and on the particular phenomena one is interested to
  study. Let us consider, for instance, the importance of inertia in
  the modelization of starling flocks
  dynamics~\cite{ACCGGJMPPSV14,CCGGJMMPSVW15}. In the present case we
  are not interested in phenomena such as the transmission of
  information across the system~\cite{ACCGGJMPPSV14,CCGGJMMPSVW15},
  for which the choice among overdamped or inertial dynamics is
  determinant, but rather on the stationary probability distribution
  of velocities in a dense state. Therefore we do not have a
  particular reason to abandon the standard fixed by the literature on
  dense glass-forming systems, i.e., overdamped dynamics, in favour of
  inertial dynamics. Moreover, in absence of confinement and on long
enough time scales, our discussion might be suitable also for inertial
active particles \cite{LeoniPaoluzzi,Lowen,CapriniMarconi}}.\\ Setting
the mobility to $\mu=1$, we consider the active dynamics characterized
by the following equations
\begin{equation} \label{eq:dynamics}
    \dot{\mathbf{x}}_i = - \nabla_i \Phi + \mathbf{f}_i
\end{equation}
with $\mathbf{f}_i$ the active force acting on each
particle. Depending on the system of interest, different prescriptions
for the dynamical evolution of the active force are possible:
Run-and-Tumble (RT) dynamics in case of swimming bacteria as
E. coli~\cite{PhysRevE.48.2553}, Active Brownian (AB) dynamics for
active colloids~\cite{romanczuk2012active} or Active
Ornstein-Uhlenbeck (AOU) dynamics, which describes well the motion of
passive objects in active baths~\cite{Bechinger17}.  Despite they are
microscopically different, all these models capture the same
phenomenology on large scales. Particularly interesting is the case of
AOU particles that admits an effective equilibrium
picture~\cite{Szamel14,Farange15,MM15,MGPMD16,Paoluzzi,PaoluzziXY,Marconi17,PhysRevResearch.2.023207}. In
AOU models, the dynamics of the self-propulsion force is an
Ornstein-Uhlenbeck process with characteristic timescale $\tau$:
\begin{equation} \label{eq:aoup}
    \dot{\mathbf{f}}_i = -\frac{1}{\tau} \mathbf{f}_i + \boldsymbol{\eta}_i
\end{equation}
where the stochastic drive $\boldsymbol{\eta}_i$ satisfies 
\begin{eqnarray}
\langle \eta_i^\alpha \rangle &=& 0 \\ 
\langle \eta_i^\alpha(t) \eta_j^\beta(s) \rangle &=& \frac{2 T}{\tau^2} \delta_{ij} \delta^{\alpha \beta} \delta(t-s). \;
\end{eqnarray}
Because of Eq.~(\ref{eq:aoup}), the self-propulsion force is actually a
coloured noise with exponential kernel:
\begin{align}
\langle f_i^\alpha(t) f_j^\beta (s)
\rangle = \frac{2 T}{\tau} \delta_{ij} \delta^{\alpha \beta} e^{-
  |t-s| / \tau }
\end{align}
The control parameters of the model are the correlation time of the
noise $\tau$, that determines the persistence time of the
self-propelled motion and its amplitude $T$. As shown in
Ref.~\cite{MGPMD16}, under suitable assumptions the stationary
probability distributions of velocities for a given steady-state
configuration of the system can be written as a Gaussian distribution
whose covariance matrix depends parametrically on the configuration:
\begin{eqnarray} \label{eq:pdv}
P(\dot{\mX}|\mX) &=& \mathcal{N} \times e^{- \frac{1}{2 T} \dot{\mX}^T \boldsymbol{\Gamma} \dot{\mX} } \\
\boldsymbol{\Gamma} &\equiv& \boldsymbol{1} + \tau \mathbb{H} \; ,
\end{eqnarray}
where ${\bf 1}$ is the $3N\times 3N$ identity matrix, $\mathcal{N}$
a normalisation constant, and $\mX \equiv (
\mathbf{x}_1,...,\mathbf{x}_N)$. In particular, given the Gaussian
form of the conditional probability distribution in
Eq.~(\ref{eq:pdv}), the correlation between the velocities of
different particles reads as
\begin{align}
  \langle \dot{x}_i^\alpha \dot{x}_j^\beta \rangle = (\boldsymbol{\Gamma}^{-1})_{ij}^{\alpha \beta}.
  \label{eq:corrv}
\end{align}
The beautiful insight suggested by Eq.~(\ref{eq:corrv}) is that such
off-equilibrium correlations are related with the non-diagonal
elements of the Hessian matrix. \gcol{It is also clear from
  Eq.~(\ref{eq:pdv}),(\ref{eq:corrv}) that in the equilibrium limit
  all such correlations become trivial, as they have to
  \begin{align}
    \lim_{\tau\rightarrow 0} \langle \dot{x}_i^\alpha \dot{x}_j^\beta \rangle \propto \delta_{ij}\delta_{\alpha\beta}.
    \label{eq:corr-equil}
  \end{align}
  The most important condition under which the conditional probability
  in Eq.~(\ref{eq:pdv}) is well defined and the expression in
  Eq.~(\ref{eq:corrv}) holds for \emph{all} values of the persistence
  time $\tau$ is to have a \emph{positive-defined} Hessian
  matrix. Something which is true, in the case of standard interaction
  potentials, for low temperature quasi-crystalline states or for
  glassy arrested ones. As soon as unstable directions arise in the
  Hessian, e.g., for typical liquid state configurations, one is
  bounded to small enough values of the persistence time $\tau$ for
  the above expression of $P(\dot{\mX}|\mX)$ to be valid. It is only
  when $\tau = 0$ that $P(\dot{\mX}|\mX)$ is always well defined,
  namely both in the arrested and in the liquid phase. But with
  $\tau=0$ velocity correlations are trivial, see
  Eq.~(\ref{eq:corr-equil}). Alternatively one should look for more
  refined, but also more difficult to derive rigorously, non-Gaussian
  forms of $P(\dot{\mX}|\mX)$, for instance the one proposed at the
  end of Sec.~\ref{sec:four}. To the purpose of the present analysis
  the coloured noise characteristic time-scale $\tau$ is much more
  relevant with respect to its amplitude, since the consistency of the
  velocity distribution $P(\dot{\mX}|\mX)$ as written in
  Eq.~(\ref{eq:pdv}) depends solely on $\tau$. This notwidthstanding
  we will play with both parameters: after having equilibrated the
  system at a given temperature, applying active noise of small
  amplitude is functional to not alter the equilibrium landscape,
  while a large enough value of $\tau$ allow to have non-trivial
  non-equilibrium correlations.}

  \gcol{The key interesting feature of the conditional probability
    $P(\dot{\mX} | \mX)$ written in Eq.~(\ref{eq:pdv}) is the
    following:} it allows to relate in a very transparent manner
  non-equilibrium correlations and the equilibrium topology of a
  complex liquid energy landscape.  \gcol{Our analysis will consist in
    studying the properties of these correlations in a phase where the
    Hessian is positive-definite, namely the glassy arrested state,
    and in drawing some conclusions on the possible behaviour in the
    liquid phase, where the amplitude of velocity fluctuations becomes
    apparently unbounded according to the expression of
    $P(\dot{\mX}|\mX)$ in Eq.~(\ref{eq:pdv}).} We will start by
  illustrating the situation in an exactly-solvable model where the
  minima-to-saddles topological transition is an analytical result in
  the large-$N$ limit.

\section{Active $p-$spin}
\label{sec:three}

\begin{figure}
  \includegraphics[width=0.9\columnwidth]{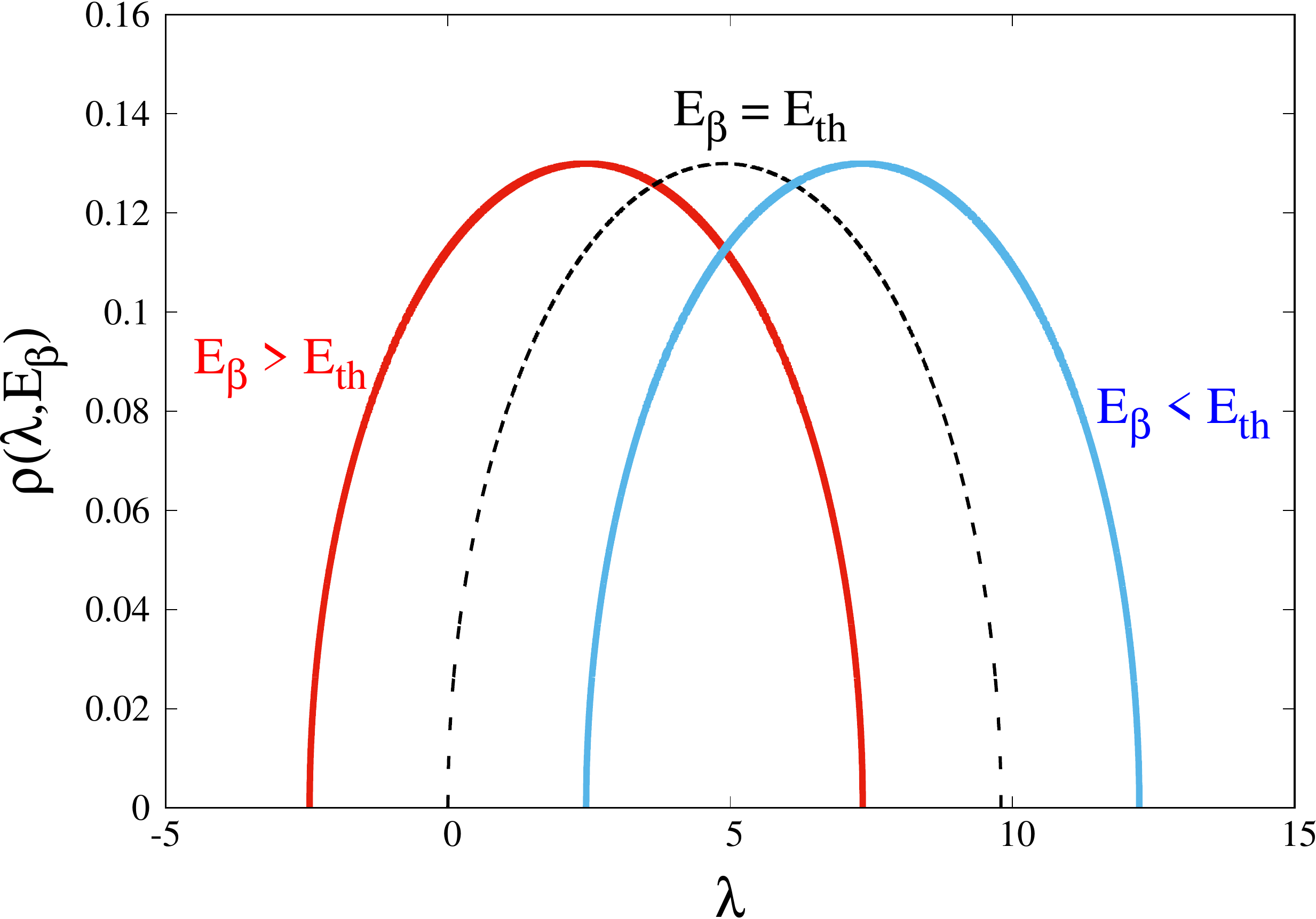}
  \caption{Average eigenvalues spectrum $\rho(\lambda, E_\beta)$ for
    the Hessian $\mathbb{H}$ of the $p$-spin model [see
      Eq.~(\ref{eq:Hess-pspin-av})], computed for different values of
    the stationary points energy $E_\beta$ and for $p=4$
    ($E_{\text{th}}=\sqrt{2(p-1)/p}$) :
    $E=E_{\text{th}}-\frac{E_{\text{th}}}{2}$ ($E>E_{\text{th}}$),
    $E=E_{\text{th}}+\frac{E_{\text{th}}}{2}$ ($E<E_{\text{th}}$) and
    $E=E_{\text{th}}$.}
\label{fig1}
\end{figure}

\gcol{The connections between off-equilibrium non-diagonal velocity
  correlations and the eigenvalues spectrum of potential energy
  Hessian are clearly illustrated by a model which can be solved
  exactly in the mean-field approximation: a non-equilibrium version
  of the disordered spherical $p$-spin model. Despite recent results
  showed that the $p$-spin model with non-homogeneous interaction
  potentials better captures some features of the aging dynamics of
  realistic glass-formers~\cite{FFT20} (e.g., polydisperse mixtures
  with Lennard-Jones potentials), we stick here to the ``traditional''
  $p$-spin with homogeneous interactions as the simplest model to
  understand the more fundamental properties of driven glassy
  systems. Since the present one is the first attempt to relate
  landscape topology to properties of correlations in a
  \emph{stationary} non-equilibrium regime, the simplest version of
  the $p$-spin model should be already good enough. A driven version
  of the homogeneous $p$-spin has been for instance considered also
  in~\cite{BK13}, though with a different driving mechanism. But,
  before dwelling on the specific non-equilibrium version of the
  model, let us first recall its equilibrium properties.} The $p$-spin
model is characterized by the following disordered Hamiltonian and
global constraint:
\begin{align}
  \mH_J[\boldsymbol{\sigma}] &= -\sum_{i_1<i_2<...<i_p} J_{i_1 i_2
    ... i_p}\sigma_{i_1} \sigma_{i_2} ...\sigma_{i_p}\nonumber
  \\ \sum_{i=1}^N\sigma_i^2 &= N, \; .
\end{align}
where the sum in $\mH_J[\boldsymbol{\sigma}]$ runs over all
independent $p$-uplets of indices, $J_{i_1 i_2 ... i_p}$ are normal
random variates with zero mean and variance $\langle J^2\rangle \sim
1/N^{p-1}$ (guaranteing energy extensivity). \gcol{It is well known
  that equilibrium configurations of the $p$-spin sampled with
  Boltzmann measure
\begin{align}
  P_J[\boldsymbol{\sigma}] = e^{-\beta
  \mH_J[\boldsymbol{\sigma}]}~\delta\left(\sum_{i=1}^N\sigma_i^2 - N \right)
  \label{eq:peq-pspin}
\end{align}
are typically close to stationary points of the energy hypersurface
and that the topology of these stationary points depends on the
temperature~\cite{C01}. In particular, for $T<T_d$ equilibrium
configurations are close to energy minima while for $T>T_d$ they are
close to saddle points. This correspondence works because in the
large-$N$ limit this model has a biunivocal correspondence between
energy levels and temperatures, due to the self-averaging property of
the energy:
%% %%
\begin{align}
\lim_{N\rightarrow\infty}E_{J,\beta} = \overline{E_{J,\beta}} = E_\beta
\end{align}
where $E_{J,\beta}=\langle H_J \rangle_\beta$ denotes the
thermodynamic average at fixed disorder while the overline denotes the
average over disorder instances, i.e., over the random couplings
$J_{i_{1}i_{2}...i_{p}}$.} It is then known that in the $p-$spin model
the Hessian at the stationary point of the energy landscape is a GOE
matrix~\cite{cavagna1998stationary,CGG00,MNSV09,MNSV11}. \gcol{From
  this property follows} that the distribution of Hessian
eigenvalues in the large-$N$ limit, $\rho(\lambda, E_{J,\beta})$, is a
self-averaging quantity and follows the Wigner semicircle law
\cite{cavagna1998stationary,CGG00,MNSV09,MNSV11}:
\begin{align}
  &\lim_{N\rightarrow\infty} \rho(\lambda, E_{J,\beta}) = \rho(\lambda, E_\beta) = \nonumber \\
   &= \frac{1}{\pi p(p-1)} \sqrt{p^2E_{\text{th}}^2 - (\lambda+p E_\beta)^2},
\label{eq:semicircle}
\end{align}
with $E_{\text{th}}=-\sqrt{2(p-1)/p}$ the energy where the topological
transition between a minima-dominated to a saddle-dominated landscape
takes place. The eigenvalues spectrum of the $p$-spin Hessian is
represented in Fig.\ref{fig1} for values of the energy $E_\beta$ at
the threshold, $E_\beta=E_{\text{th}}$ (black dotted line), values
below the threshold, $E_\beta<E_{\text{th}}$ (continuous blue line,
right) and values above the threshold, $E_\beta>E_{\text{th}}$
(continuous red line, left).\\ \\

%%{\bf JACK WAS HERE}\\ \\

\gcol{From Eq.~(\ref{eq:semicircle}) and Fig.~\ref{fig1} it is clear
  that all eigenvalues are positive as long as energy is below the
  threshold, $E_\beta<E_{\text{th}}$~\cite{cavagna1998stationary}. By
  increasing $E_\beta$ the eigenvalue distribution shifts to the left
  and as soon as $E_\beta = E_{\text{th}}$ negative eigenvalues
  appear. We show here how the appearing of unstable directions in the
  Hessian at $E_{\text{th}}$ directly affects the behaviour of
  stationary non-equilibrium velocity correlations by considering the
  following protocol. We define a Langevin dynamics characterized by
  the presence of a persistent noise on the spins:}
\begin{align}
  \dot{\sigma}_i(t) = - \mu \sigma_i - \frac{\partial \mH_J}{\partial \sigma_i} + f_i,
  \label{eq:lang-spins-1}
\end{align}
where the Lagrange multiplier $\mu$ enforces the spherical constraint
and the active driving force $f_i$ evolves according to an
Ornstein-Uhlenbeck process
\begin{align}
  \dot{f}_i(t) = - \frac{1}{\tau}f_i + \eta_i.
  \label{eq:lang-spins-2}
\end{align}
\gcol{First, it is convenient to set $\tau=0$ and sample the
  equilibrium distribution $\exp(-\beta\mH_J[{\bf \sigma}])$. Then,
  one switches on ``memory'' in the noise, i.e., sets $\tau>0$, and
  allows the persistent dynamics to run
  [Eqns.~(\ref{eq:lang-spins-1}),(\ref{eq:lang-spins-2})] until when a
  stationary distribution is reached. As long as the energy of the
  initial configuration is below the threshold, i.e., $E_\beta <
  E_{\text{th}}$, active dynamics stays close to the bottom of an
  energy minimum: in the $p$-spin model barriers are extensive and
  dynamical fluctuations, either thermal or active, cannot help to
  jump barriers in the thermodynamic limit.}

\gcol{Let $P_J(\boldsymbol{\sigma})$ be the equilibrium distribution
  probed by the Langevin dynamics when $\tau=0$, namely the one
  written in Eq.~(\ref{eq:peq-pspin}). The above protocol of drawing
  equilibrium configurations first and then run the active dynamics
  amounts to assume that the stationary joint distribution of
  $\boldsymbol{\sigma}$ and $\dot{\boldsymbol{\sigma}}$, consistently
  with the Unified Colored Noise
  approximation~\cite{Hanggi95,MMGD15,MGPMD16,Marconi17}, reads simply
  as:}
\begin{align}
P_J(\boldsymbol{\sigma}, \dot{\boldsymbol{\sigma}}) = P_J^{(\tau)}(\dot{\boldsymbol{\sigma}}
|\boldsymbol{\sigma}) P_J(\boldsymbol{\sigma})
\end{align}
where the conditional probability distribution of the velocities reads
as
\begin{align} \label{eq:pdv_pspin}
  P_J(\dot{\boldsymbol{\sigma}} |\boldsymbol{\sigma}) &=
  \mathcal{N} \times e^{- \frac{1}{2 T} \dot{\boldsymbol{\sigma}}^T \boldsymbol{\Gamma}_J(\boldsymbol{\sigma}) \dot{\boldsymbol{\sigma}} } \nonumber \\
  \boldsymbol{\Gamma}_J(\boldsymbol{\sigma}) &
  \equiv \boldsymbol{1} + \tau~\mathbb{H}_{J,\boldsymbol{\sigma}} \;.
\end{align}
In the present case the covariance matrix
$\boldsymbol{\Gamma}_J(\boldsymbol{\sigma})$ depends both on the
particular configuration with respect to the Hessian is
calculated and on the realization of the quenched random
couplings. The Hessian matrix is defined as:
\begin{align}
  \left[\mathbb{H}_{J,\boldsymbol{\sigma}}\right]_{i,j} =
  \frac{\partial^2\mH_{J}(\boldsymbol{\sigma})}{\partial\sigma_i\partial\sigma_j},
\end{align}
so that velocities cross correlations are related to the non-diagonal
elements of the Hessian:
\begin{align}
\mathbb{E}_{J,\boldsymbol{\sigma}}\left[ \dot{\sigma}_i\dot{\sigma}_j \right] = [\Gamma_J^{-1}(\boldsymbol{\sigma})]_{ij}, 
\end{align}
where
\begin{align}
  \mathbb{E}_{J,\boldsymbol{\sigma}}\left[ \dot{\sigma}_i\dot{\sigma}_j \right] =
  \int \mD\dot{\sigma}~P_J(\dot{\boldsymbol{\sigma}} |\boldsymbol{\sigma})~\dot{\sigma}_i\dot{\sigma}_j .
\end{align}
The last step is to consider the large-$N$ limit,
where, due to self-averaging, it is possible to write
\begin{align}
  \lim_{N\rightarrow\infty}\mathbb{E}_{J,\boldsymbol{\sigma}}\left[ \dot{\sigma}_i\dot{\sigma}_j \right] & =
  \lim_{N\rightarrow\infty}\overline{\langle\mathbb{E}_{J,\boldsymbol{\sigma}}\left[ \dot{\sigma}_i\dot{\sigma}_j \right]\rangle}
  \nonumber \\
  & = \left[(\mathbb{I} + \tau \mathbb{H} )^{-1}\right]_{ij},
\label{eq:Hess-inv-pspin}
\end{align}
with
\begin{align}
  \mathbb{H} = \overline{\langle  \mathbb{H}_{J,\boldsymbol{\sigma}}\rangle},
  \label{eq:Hess-pspin-av}
  \end{align}
\gcol{and where we have denoted with $\overline{[\#]}$ the average
  over quenched randomness and with $\langle \# \rangle$ the average
  over the reference configuration. For the $p-$spin model the matrix
  $\mathbb{H}_{J,\boldsymbol{\sigma}}$ belongs to the GOE
  ensemble~\cite{CGG00,MNSV09,MNSV11}, so that in the large-$N$ limit
  we have precisely the identity written in
  Eq.~(\ref{eq:Hess-pspin-av}) and the spectrum of eigenvalues
  described by the expression in Eq.~(\ref{eq:semicircle}). From
  Eq.~(\ref{eq:Hess-inv-pspin}) we find that, as expected, in the
  equilibrium limit $\tau\rightarrow 0$ the non-diagonal correlations
  among spin \emph{``velocities''} vanish:
\begin{align}
\lim_{\tau\rightarrow\infty} \lim_{N\rightarrow\infty} \overline{\langle\mathbb{E}_{J,\boldsymbol{\sigma}}\left[ \dot{\sigma}_i\dot{\sigma}_j \right]\rangle} = \delta_{ij}.
\end{align}
Then, in order to single out the behaviour of cross correlations when
energy reaches the threshold value for the topological transition,
i.e., when $E = E_{\text{th}}-\delta E$ and $\delta E\rightarrow 0$,
it is convenient to consider the orthogonal transformation
$U:\mathbb{R}^N\rightarrow\mathbb{R}^N$ which diagonalizes $\mathbb{H}$. In the
same limit where the identity of Eq.~(\ref{eq:Hess-inv-pspin}) holds
we can write the probability distribution of velocities as:}
\begin{align}
\lim_{N\rightarrow\infty} P_J(\dot{\boldsymbol{\sigma}}
|\boldsymbol{\sigma}) = P_J^{(\infty)}(\dot{\boldsymbol{\sigma}})
\propto e^{- \frac{1}{2 T} \dot{\boldsymbol{\sigma}}^T
  [{\bf 1}+\tau \mathbb{H}] \dot{\boldsymbol{\sigma}} }.
\end{align}
Writing the Hessian in its diagonal form,
we can write the above distribution as
\begin{align}
  P_J^{(\infty)}(\dot{\boldsymbol{\sigma}}) \propto \exp\left\lbrace - \frac{1}{2 T}
  \sum_{k=1}^N |\dot{\sigma}_{k}|^2 [1+\tau \lambda_k]\right\rbrace,
  \label{eq:diagonal-P}
\end{align}
where $\dot{\sigma}_k = \sum_{i=1}^N U_{ki} \dot{\sigma}_i$.  From
Eq.~(\ref{eq:diagonal-P}) a Gaussian integration leads to
\begin{align}
\mathbb{E}[\dot{\sigma}_{k}\dot{\sigma}_{-k}] = \frac{1}{1+\tau \lambda_k},
\end{align}
from which it is easy obtained the divergence of cross-correlations
among spin velocities when the lower band edge of the spectrum cross
the origin, taken a not vanishing value of $\tau$.

\section{The signature of criticality at $T_d$ for active particles}
\label{sec:four}

It is quite reasonable to believe that the mean-field picture
introduced in the previous section holds in first approximation even
for the finite-dimensional system of interacting active particles
described by Eq.~(\ref{eq:dynamics}) in Sec.~\ref{sec:one}. \gcol{We
  again consider the protocol according to which the} configurations
$\mX$ are first sampled according to the equilibrium Boltzmann
distribution $e^{-\beta \Phi(\mX)}$ and then self-propelled motion is
switched on with parameters $\tau$ and $T=\beta^{-1}$. We expect that
the only relevant difference with the protocol outlined for the
$p$-spin model is that now, since different minima of the potential
might have finite barriers (particular close to $E_{\text{th}}$), one
should consider a not \gcol{too} large amplitude $T$ of the
noise. \gcol{A small value of $T$ does not limit the scope of} our
analysis, since the noise amplitude only affects the amplitude of
non-equilibrium correlations, not their range. \gcol{The sampling of
equilibrium configurations below $T_d$ can be easily obtained using an on-purpose smart Monte-Carlo algorithm first proposed
in~\cite{GP01} and recently brought to the top of
efficiency~\cite{NBC17,BCCNOY17}. From Eq.~(\ref{eq:pdv}) it follows
that at thermodynamic equilibrium, namely when $\tau=0$, one has:
\begin{align}
  P_{eq}(\dot{\mathbf{x}}|\mathbf{x}) \propto e^{-\frac{1}{2T} \dot{\mathbf{x}}^2}  ~~~ \Longrightarrow ~~~
    \langle \dot{{\bf x}}_i \dot{{\bf x}}_j\rangle =0~~~\forall~i\neq j,
\end{align}
whereas the possibility of having off-diagonal velocity correlations
$\langle \dot{{\bf x}}_i \dot{{\bf x}}_j\rangle \neq 0$ (for $i\neq
j$) is only realized in the presence of active dynamics, $\tau>0$, as
it was shown even in the case of granular
gases~\cite{GSVP11,GSVP11b,PGGSV12,caprini2020hidden,PhysRevLett.124.078001}. Here
we stress that, thanks to Eq.~(\ref{eq:pdv}), such non-equilibrium
correlations also reveal the properties of the energy landscape of the
system of amorphous materials as soon as some activity is switched
on.}\\ \\

%{\bf JACK WAS HERE}\\ \\ 

\gcol{Taking inspiration from the properties of the $p$-spin model in
  Sec.~\ref{sec:three}, we assume that non-diagonal velocity
  correlations are \emph{self-averaging}. In the present case there is
  no quenched disorder and the average is only over configurations
  $\mX^*$ close to stationary points of the potential energy
  landscape. The request of self-averaging for velocity correlations
  corresponds to the request of the same property for Hessian
  eigenvalues, which in the large-$N$ limit are expected to behave
  similarly in mean-field and finite-dimensional models. Concerning
  this, let us stress that the information needed to describe in first
  approximation the correlations of the velocity field is only related
  to Hessian eigenvalues and not to the corresponding eigenvectors,
  which close to transitions in the landscape may show non-trivial
  localization properties~\cite{CNB19,SCMI21}. We thus write the
  velocity correlations in the large-$N$ limit as:
\begin{align}
\lim_{N\rightarrow\infty} \langle \dot{\bf x}_i^\alpha\dot{\bf x}_j^\beta\rangle_{_{\mathcal{X}^{*}}} = \left[(\mathbb{I} + \tau
  \mathbb{H} )^{-1}\right]_{ij}^{\alpha\beta},
\label{eq:Hess-inv-offlattice}
\end{align}
where
\begin{equation}
  \mathbb{H} = \langle \mathbb{H}\rangle_{\mX^*}
\end{equation}
}

Considering the orthogonal matrix
$U:\mathbb{R}^{3N}\rightarrow\mathbb{R}^{3N}$ which diagonalizes the
Hessian we can then change variables to
\be {\bf u}_q = \sum_{i=1}^N U_{qi} ~\dot{{\bf x}}_i. \ee
In terms of the new rotated variables the probability distribution of
velocities reads as
\begin{align}
 \lim_{N\rightarrow\infty} P({\bf U}|\mX^*) \propto
e^{-\frac{1}{2 T} \sum_{q=1}^{3N} (1+\tau \lambda_q) |{\bf u}_q|^2},
\label{eq:diag-prob}
\end{align}
where ${\bf U}=({\bf u}_1,\ldots,{\bf u}_N)$, $\lambda_q$ are the eigenvalues of $\mathbb{H}$ and the
amplitude of the $q$-th mode reads as
\be \langle |{\bf u}_q|^2
\rangle \sim \frac{1}{1+\tau \lambda_q}. 
\label{eq:amp-mode}
\ee
\gcol{The dependence on the configuration $\mX^*$ is lost on the right-hand
member of Eq.~(\ref{eq:diag-prob}) due to the claimed self-averaging
property. It is clear that also in this case as soon as one starts to
have negative eigenvalues of the Hessian the approximation leading to
Eq.~(\ref{eq:amp-mode}) breaks down: the occurrence of an interesting phenomenon is signaled by the divergence of velocity correlations.}\\

A similar picture is supported by the following argument (see
  Refs. \cite{henkes2020dense,PhysRevE.84.040301,PhysRevX.6.021011}
  for details). Let $\mX^*$ be a stationary point of the potential
energy landscape, i.e., $ \left. \nabla \Phi \right|_{\mX^*}=0$.  We
indicate with $\delta \mX = \mX - \mX^*$ a small fluctuation around
the inherent state. The (linearized) equations of motion for the
fluctuations are
\begin{equation} \label{eq:fluct}
    \delta \dot{\mathbf{x}}_i = - \sum_{j=1}^N \mathbb{H}_{ij}(\mX^*)\cdot \delta \mathbf{x}_j + \mathbf{f}_i\; ,
\end{equation}
where $\mathbb{H}_{ij}$ indicates the dynamical matrix, i.e., the
Hessian computed at $\mX^*$. In particular $\mathbb{H}_{ij}$ denotes a $d \times d$
block, so that \gcol{Eq.~(\ref{eq:fluct}) must be regarded as a compact notation for a vectorial equation
of the kind:
\begin{equation} \label{eq:fluct2}
  \delta \dot{x}_i^\alpha = - \sum_{j=1}^N \left[ \mathbb{H}_{ij}(\mX^*)\cdot \delta \mathbf{x}_j \right]^\alpha
  + f_i^\alpha \; ,
\end{equation}
where}
\begin{align}
  \left[\mathbb{H}_{ij}\cdot \delta \mathbf{x}_j\right]^\alpha =
  \sum_{\beta=1}^d \mathbb{H}_{ij}^{\alpha\beta}~\delta x_j^\beta 
\end{align}

In
writing Eq.~(\ref{eq:fluct}) we consider that fluctuations are driven
by an active dynamics, see Eq.~(\ref{eq:dynamics}). For concreteness,
let us consider the case of a 2d-system with AB dynamics:
\begin{align}
  \mathbf{f}_i &= v_0 (\cos \theta_i, \sin \theta_i), \nonumber \\
  \dot{\theta}_i &=\eta_i,
\end{align}
where $\langle \eta_i \rangle = 0$ and
\begin{align}
\langle \eta_i(t) \eta_j(s) \rangle &= 2 \tau^{-1} \delta_{ij} \delta(t - s) 
\end{align}
By expading the fluctuations on a normal modes basis 
\begin{align}
\delta \mathbf{x}_i(t) = \sum_{{\bf q}} a_{\bf q}(t) \mathbf{u}_i({\bf q}),
\end{align}
one obtains that the amplitudes $a_{\bf q}(t)$ evolve
according to \cite{henkes2020dense}
\begin{equation}
\dot{a}_{\bf q}(t) = -\lambda_{\bf q} a_{\bf q}(t) + \tilde{\eta}_{\bf q}(t),    
\end{equation}
with an exponentially correlated noise
\begin{align}
\langle \tilde{\eta}_{\bf q}(t) \tilde{\eta}_{\bf p} (s) \rangle = \frac{v_0^2}{2}
\,\delta_{{\bf q},{\bf p}}~e^{-|t-s|/\tau } \; .
\end{align}
After standard manipulations
\matteo{\cite{henkes2020dense}}, one obtains
\begin{align}
  \label{eq:power-spectrum}
  \langle \dot{a}_{\bf q}^2\rangle \sim (1 + \tau \lambda_{\bf q} )^{-1},
\end{align}
that is again in agreement with Eq.~(\ref{eq:amp-mode}).  Clearly as
long as \gcol{the active dynamics keeps the system in the close
  vicinity of stationary points of the minima-dominated landscape the
Hessian, which is computed with respect to these stationary points, is
a positive definite matrix. Of course, despite the configurations of
the system are close to equilibrium, i.e. close enough for the
harmonic approximation of Eq.~(\ref{eq:fluct}) to be valid, the
velocity field has typical non-equilibrium features. In particular, by
Fourier antitrasforming the power spectrum in
Eq.~(\ref{eq:power-spectrum}), it is immediate to realize that the
velocity amplitudes correlation function is nontrivial, whatever the
shape of the spectrum $\lambda_{\bf q}$:
\begin{align}
\langle a({\bf x}) a({\bf y}) \rangle \neq \delta({\bf x}-{\bf y}).
\end{align}
On the contrary, in the presence of thermodynamic equilibrium, i.e.,
when $\tau= 0$, independently to the amplitude of noise one always
has:
\begin{align}
  \langle \dot{a}_{\bf q}^2\rangle ~ = ~ \textrm{const}~~~\Longrightarrow~~~\langle a({\bf x}) a({\bf 0}) \rangle \propto \delta({\bf x}).
\end{align}
Therefore, while at equilibrium the matrix $({\bf 1} + \tau
\mathbb{H})$ is simply diagonal and does not contain any information
on the landscape, close to crystalline or quasi-crystalline or glassy
arrested states it has a non-trivial structure. And, something that is
of particular importance for the present analysis, it is positive
definite for \emph{any} value of $\tau$. From Eq.~(\ref{eq:amp-mode})
it is immediate to see that in this case, i.e. for energies
$E<E_{\text{th}}$, the power spectrum of velocity modes is well
defined for any value of ${\bf q}$, since both $\tau$ and
$\lambda_{\bf q}$ are positive. On the contrary, as soon as some
eigenvalues become negative, one can always find a value of $\tau$ for
which the expression $\langle |{\bf u}_{\bf q}|^2 \rangle \sim (1-\tau
|\lambda_{\bf q}|)^{-1}$ is inconsistent. Physically, we can relate
this instability to the minima-to-saddles crossover. The apparent
divergence of velocity correlation function there occurring might
signal the presence of a non-equilibrium transition between a
phase where the velocity field is disordered, for temperatures $T<T_d$,
and a high temperature phase at $T>T_d$ where, due to the combined
effect of the persistent noise and the non-trivial interaction
potential, some sort of order arises. It is necessary to go beyond
a Gaussian Ansatz to find an expression for $P(\dot{\mX}|\mX)$ which
remains consistent at the topological crossover and at higher
temperatures. Considering the symmetry of the system, the simplest
choice for stabilizing the velocity distribution along the unstable
directions could be}

\begin{equation}
P^{(\infty)}({\bf U}) \sim \exp{ \left\{ -\frac{1}{2T}  (1 + \tau \lambda_\kappa) |\mathbf{u}_\kappa|^2 + \frac{b}{4}  |\mathbf{u}_\kappa |^4 \right\} }
\end{equation}
with $b>0$ a positive constant. 
It is worth noting that a systematic study might be done considering higher-order terms in the
expansion of Eq.~(\ref{eq:fluct}).\\

\gcol{Let us conclude this analysis with a summary of the
  physical role of the two parameters of the coloured noise: its
  characteristic time-scale $\tau$ and its amplitude $v_0$. The
  smallness of $v_0$ is what allows us to consider simply a harmonic
  expansion of the potential close to a stationary point and plug it into
  the study of active off-equilibrium correlations information
  which is basically the one on equilibrium stationary configurations.
  On the contrary, the presence of $\tau$ gives rise to non-diagonal
  velocity correlations which are totally absent at equilibrium. Such
  velocity correlations might even have a small \emph{amplitude}, if
  $v_0$ is small, by in the presence of a finite $\tau$ have a finite
  range (something which would not possible at equilibrium), a range
  that we expect to become critical at the topological
  transition. That is how a combination of small $v_0$ and finite
  $\tau$ reveals the topological properties of the landscape.}

\section{Discussion and Conclusions}
\label{sec:conclusions}
In the present paper, we have proposed a theoretical insight that
connects the topological crossover in glasses with non-equilibrium
velocity correlations that are typical of dense active matter systems
\cite{henkes2020dense,PhysRevLett.124.078001,caprini2020hidden,caprini2020active}.
\gcol{We showed that, to the extent of a Gaussian approximation for
  the marginal joint distribution of velocities, a blow-up of velocity
  fluctuations takes place at the saddle-to-minima topological
  crossover in glass-forming systems. This blow-up can be regarded as
  the likely signature of a non-equilibrium phase transition.}  This
scenario is suggested both by the trial distribution for the
velocities suggested in Ref.~\cite{MGPMD16} and by the one drawn from
the UCNA approximation~\cite{MM15}. Moreover, the same scenario
emerges performing the linear stability analysis around a stationary
point of the potential energy landscape \cite{henkes2020dense}.  Our
analysis shows that, around the minima-to-saddle crossover,
off-diagonal correlations due to the self-propulsion make the
inherent configuration unstable. Moreover, at the crossover, velocity
fluctuations tend to diverge. \gcol{Our analysis suggests that while
  no long-range order in the velocity field takes place for active
  glassy states below $E_{\text{th}}$, at the minima-to-saddles there
  are signatures of something non-trivial occurring, perhaps a
  crossover to a flocking phase or the formation of living
  crystals~\cite{B13}. It is worth noting that in the systems
  considered the transition is not triggered by an alignment
  interaction but is solely due to the combined effect of persistent
  noise and a non-trivial interaction potential, a scenario compatible
  with recent results~\cite{PhysRevLett.124.078001,caprini2020hidden}
  and which is surely worth to investigate with more detailed
  numerical simulations in the near future.}\\

\section*{Acknowledgments.}
G.G. and M.P. acknowledge J.-L. Barrat, E. Bertin, L. Caprini,
A. Cavagna, S. N. Majumdar, V. Ros and G. Sicuro for useful
conversations and the Physics Department of ``Sapienza'', University
of Rome, for kind hospitality at some stages during this manuscript
preparation. M.P. has received funding from the European Union's
Horizon 2020 research and innovation programme under the MSCA grant
agreement No 801370 and by the Secretary of Universities and Research
of the Government of Catalonia through Beatriu de Pin\'os program
Grant No. BP 00088 (2018).

\end{document}